# Superconducting nanowire single-photon detectors at a wavelength of 940 nm

W. J. Zhang, H. Li, L. X. You[*], Y. H. He, L. Zhang , X.Y. Liu, X. Y. Yang, J. J. Wu, Q. Guo, S. J. Chen, Z. Wang and X. M. Xie

*State Key Laboratory of Functional Materials for Informatics, Shanghai Institute of Microsystem and Information Technology (SIMIT), Chinese Academy of Sciences (CAS), 865 Changning Rd., Shanghai 200050, China.*

[*]*E-mail: lxyou@mail.sim.ac.cn*

## ABSTRACT

We develop single-photon detectors comprising single-mode fiber-coupled superconducting nanowires, with high system detection efficiencies at a wavelength of 940 nm. The detector comprises a 6.5-nm-thick, 110-nm-wide NbN nanowire meander fabricated onto a Si substrate with a distributed Bragg reflector for enhancing the optical absorptance. We demonstrate that, via the design of a low filling factor (1/3) and active area ($\Phi = 10$ μm), the system reaches a detection efficiency of ~60% with a dark count rate of 10 Hz, a recovery time <12 ns, and a timing jitter of ~50 ps.

## MAIN TEXT

The efficient generation and detection of single photons lies at the heart of quantum information (QI) processing[1-4]. Self-assembled InAs/GaAs quantum dots (QDs) have proven to be an efficient source of single photons with high purity (99.7%) and indistinguishability (99.5%)[5]. They hold the promise of offering an excellent solid-state platform for investigations of multiphoton entanglement[1], boson sampling[6], and linear optical quantum computing[7]. However, a major experimental challenge[4] is the lack of high-performance detectors at the emission wavelength range of these QDs, which is typically around a wavelength of 940 nm. Silicon avalanche photodiodes (APDs) have low efficiencies (20%−30%) and long dead times (~45 ns) at a wavelength of 940 nm[5,8], with inevitable afterpulsing counts. Thus, the development of faster, more efficient single-photon detectors at a wavelength of 940 nm is critical to the implementation of QI technologies in the future.

Superconducting nanowire single-photon detectors (SNSPDs) have proven to be promising detectors for near infrared photons because of their high detection efficiency (>90%), low dark count rate (<100 Hz), low timing jitter (<70 ps), and short recovery time (<40 ns). For instance, WSi-SNSPD demonstrated a record system detection efficiency (SDE) of 93% at a wavelength of 1550 nm[9], though it needs a sub-Kelvin operating temperature. Nb(Ti)N-SNSPDs have demonstrated an SDE of over 75% at a moderate temperature over 2 K[10,11]. Waveguide-coupled SNSPDs have shown a fast-recovery time (<10 ns), but the SDE is still limited to <20%[12,13]. Indeed, SNSPDs may provide excellent performance not only at 1550 nm but also at shorter wavelengths, since photon energy scales as the inverse of wavelength. Recently, Liu et al demonstrated multimode fiber-coupled SNSPDs with SDEs of over 70% at a wavelength of 635 nm[14]. However, there has been little research on developing high-performance SNSPDs at wavelengths around 940 nm, though there were some reports on the characterization of QD single-photon sources using SNSPDs with relatively poor performance at these wavelengths[3,8,13]

In this letter, we report single-mode fiber-coupled SNSPDs at a wavelength of 940 nm with high SDE and fast recovery time. The detectors were fabricated on a Si substrate with a multilayer dielectric mirror to realize high optical absorptance with front-side illumination. By optimizing the design and the fabrication for photons with a wavelength of 940 nm, we demonstrate that SNSPDs can achieve an SDE of ~60% with a dark count rate (DCR) of 10 Hz, a recovery time <12 ns, and timing jitter of ~50 ps.

The optical absorption of the nanowire is one of the key parameters to consider for achieving a high SDE. Optical cavity structures on a Si substrate have been demonstrated to be an effective way to enhance the optical absorption of SNSPDs at a wavelength of 1550 nm[10,11]. However, the structure is not applicable for wavelengths below 1.2 μm since the photons are reflected and absorbed by the Si substrate. Consequently, another structure with a distributed Bragg reflector (DBR) on a Si[14,15] or GaAs[16] substrate was developed for SNSPD instead, with which high absorption can be guaranteed with front-side illumination. This structure also provided a high flexibility, which was independent of the substrate material, for various target wavelengths as long as the thicknesses of the DBR films were changed.

Figure 1(a) illustrates the simulation results for the reflectance ($R$) of the DBR substrate without NbN thin film and for the absorption efficiency ($A$) of SNSPD of two polarization states as functions of wavelength. The full detector structure from top to bottom was NbN/DBR/Si, as shown in Fig. 1(b). The simulations were based on nanowires with a linewidth/spacing of 110/220 nm and a thickness of 6.5 nm by using the RF module of COMSOL multiphysics[17]. The simulated $A$ showed a broadband enhancement around a wavelength of 940 nm. The absorptions in the parallel ($A_{//}$) and perpendicular ($A_{\perp}$) polarization states reached 94% and 49%, respectively, resulting in a PER (polarization extinction ratio, $A_{//}/A_{\perp}$) of 1.9.

The DBR substrate was a 13 periodic $Ta_2O_5/SiO_2$ bi-layers deposited on a silicon wafer with 99% reflectance measured at a wavelength of 940 nm. 6.5-nm-thick NbN films were grown on DBR substrate by direct-current magneto-sputtering. The root-mean-square roughness of NbN thin films on DBR substrate were 0.18 nm, comparable to the ones on thermally oxidized silicon substrates. The meander nanowires were patterned into NbN films by electron-beam-lithography and reactive ion etching using $CF_4$ plasma. Aiming at QI applications at a wavelength of 940 nm, which need high SDE as well as fast recovery times, the nanowire was designed to have a low filling factor of 1/3 with a linewidth/spacing of 110/220 nm and a circular active area. SNSPDs of different active area diameters of $\Phi$ = 5–18 μm were fabricated on the same wafer. In this way, we could study the SDE relations on the optical coupling and the size dependence of the recovery time. Figure 1(c) shows the scanning electron micrograph of SNSPD with a diameter of 10 μm. The well-defined nanowire, shown in Fig. 1(d), guaranteed a high uniformity of SNSPD.

The devices were mounted on a fiber-coupled package optimized for front-side illumination. A 940-nm single-mode fiber (Thorlabs Inc., SM-800) with a core diameter of 5.6 μm was adopted for optical coupling. The packages were then installed in a cryostat based on a Gifford–McMahon cryo-cooler, which had a minimal temperature of 2.2 K. To optically probe the devices, a 940-nm continuous-wave fiber laser and variable attenuators were adopted. A fiber polarization controller was inserted after the attenuators to change the polarization of the photons. Since there are no commercially available variable attenuators for 940 nm light, we carefully calibrated the attenuators manually[18], so that the light was attenuated to a typical photon flux of 0.1 M/s. The fiber was cut and spliced with the one attached to the detector in the cryostat. The photon response signal from SNSPD was amplified by a room temperature low-noise amplifier (RF Bay Inc., LNA650), and then counted using a pulse counter (SRS Inc., SR400).

The statistics of 50 devices indicated that SNSPDs had a critical temperature, $T_C \sim 7.6 \pm 0.4$ K. We measured the switching current ($I_{sw}$) of the devices with different diameters (10 devices for each size). The results indicated that the $I_{sw}$s ranged from 14.6 to 24.0 μA. Considering the wire width ($w$) ~ 110 nm, and thickness ($t$) ~ 6.5 nm, we obtained a switching current density $J_{sw} = I_{sw}/w/t = 2.0$–$3.4$ MA/cm$^2$, i.e. $2.7 \pm 0.7$ MA/cm$^2$. Figure 2(a) shows the highest SDEs for various active areas ($\Phi = 5, 10, 15$, and $18$ μm) as functions of the bias current ($I_b$) for the parallel polarization. All the SDEs were saturated with the increase of the bias current, indicating high intrinsic detection efficiencies (IDE) of all the detectors. On the other hand, the SDEs increased with the increase of the active area and saturated for $\Phi \geq 15$ μm, indicating an excellent optical coupling efficiency (OCE) for large active area SNSPDs. The highest SDE reached 66.5% ±2.5% for $\Phi = 18$ μm. $I_{sw}$ decreased as $\Phi$ increased, probably due to the increasing defects in nanowires. However, the SDEs were not influenced by the defects, because of the broad saturated plateau in the SDE–$I_b$ curves. Moreover, the highest SDEs can be reached even at a low DCR of ~ 10 Hz indicated by the dashed line in Fig. 2(b). The SDEs included all losses in the system and the overall relative errors of the SDE values could be estimated to be 3.7%, resulting from the relative uncertainty of the power meter calibration (3.50%), the long-term stability of the light power (1.14%), and attenuator calibration (0.55%).

Figure 2(c) shows the wavelength dependence of SDE in parallel polarization (SDE$_{//}$) *for the SNSPD with $\Phi = 15$ μm*. SDE$_{//}$ of 19.2% ±0.7%, 65.4% ±2.4%, and 51.1% ±1.9% were obtained for wavelengths of 850, 940, and 1064 nm, respectively, which is qualitatively consistent with the simulation results in Figure 1(a). We observed saturated plateaus in SDE for all these wavelengths. As the wavelengths increased, the saturation plateau turned narrower and the transition range became wider. The SDE$_\perp$ for 940 nm with perpendicular polarization was also measured and shown in Figure 2(c), which gave the PER value of about 1.8, slightly smaller than the simulation result, probably due to the geometric parameter deviations of the practical devices with the design parameters.

Next, the counting rate (CR) and the recovery time of the SNSPDs were investigated. By increasing the photon intensity of the illumination, we obtained the CR dependence of the SDE at 0.90 $I_{sw}$ (Figure 3(a)). When CR varied from 0.1 to 1 MHz, the SDEs were almost constant. As the CR increased, the detector could not respond to all the incoming photons limited by the slow current recovery process of the SNSPD, resulting in a decrease in the measured SDE. By defining $CR_{exp}$ as the SDE with the 1/e suppression from its maximum value, the $CR_{exp}$ values were found to be 22 ±2, 34 ±3, 68 ±5, and 72 ±6 MHz for $\Phi$ = 18, 15, 10, and 5, respectively. Figure 3(b) shows the averaged temporal response pulses recorded by a high-speed, real-time oscilloscope (Tek DSA 71254). The recovery times ($\tau_r$), defined as the time required for the pulse to decay from 90% to 10% of the maximum of the pulse, were 44.4, 26.2, 11.9 and 4.8 ns, for $\Phi$ = 18, 15, 10, and 5 μm, respectively.

Correspondingly, $CR_r = 1/\tau_r$ were 23, 38, 84, and 208 MHz.[19] In general, small active areas were evidently beneficial in obtaining a high CR. However, we noticed that $CR_{exp}$ for SNSND with $\Phi \leq 10$ μm ($L_k \leq 345$ nH[20]) was not as high as the expectation $CR_r$, due to the increase of the latching possibility under bright light illumination. This similar latching effect was reported in low $L_k$ devices on SiO$_2$/Si substrates[10,19]. Thus, a further increase in CR by reducing the active area or $L_k$ could be stemmed by the latching effect.

The timing jitter of the SNSPD was also evaluated using a time-correlated single-photon counting (TCSPC) module and 100-fs pulsed laser at a wavelength of 1550 nm and a repetition rate of 20 MHz[21]. The intensity of the laser was set to be 0.1 photon/pulse to assure the SNSPD was in single-photon-response regime. The timing jitter defined by the full width at half maximum of the histogram was 49.6 ps at the bias current of 18.0 μA (0.90 $I_{sw}$) for the SNSPD with $\Phi = 10$ μm. For all the detectors we measured, the timing jitter ranged in value from 40 to 60 ps, depending on the signal-to-noise ratio of the voltage pulse. It is noted here that the timing jitter measured at 1550 nm could be different from the value measured at 940 nm due to the different mechanisms (hot-spot or vortex-based mechanisms)[22].

SDE can be expressed as SDE = IDE × OCE × AE[9], where AE is the absorption efficiency. The experimental results above indicated that IDE and OCE (for $\Phi \geq 15$ μm) are nearly 100%; thus, we could estimate that AE ~ SDE ~ 67%. The reasons for the deviation of $AE$ with $A_{//}$ (94%) are complicated. For example, a non-ideal DBR structure may cause a lower $A_{//}$ compared with the calculated value. Geometric parameters of the practical nanowire also deviated from the designed values due to the nanofabrication process. Besides, the calculations were all based on the reflective indices of the materials obtained at room temperature. Different reflective indices at low temperature could also effectively change the final AE[23,24]. Indeed, both experiments and calculations have proven that SNSPD with a moderately high filling factor will give a high SDE[25,26]. The SDEs of ~60% (the maximal value is ~67% for $\Phi = 18$ μm) were demonstrated in our study with a low filling factor of 1/3 which is a trade-off between the high SDE and the short recovery time[25]. The maximal SDE is slightly lower than the values reported in other literatures for other wavelengths (>75% at 1550 nm[9,11] and >70% at 635 nm[14].) where the SNSPDs with a high filling factor (~0.6) were studied. On the other hand, the SDE of 67% is comparable with the previous results (<70%) for the SNSPDs with the same filling factor of 1/3 at a 1550-nm wavelength[25]. Furthermore, a low filling factor nanowire is beneficial to reduce the current crowding effect[26,27], giving a high yield and broaden the saturated plateau in the SDE–$I_b$ curves (see Fig. 2). To further enhancing the CR, there are some useful methods, such as parallel nanowires[28], a DC-coupled cryogenic circuit[19], or increased load impedance, $R_L$[2,29]. However, the intrinsic solution should focus on decreasing the thermal relaxation time between SNSPD and the substrate to avoid latching.

In conclusion, we report the fabrication and characterization of SNSPDs at a wavelength of 940 nm. These SNSPDs showed high detection efficiencies (~60%) and fast-recovery times (~12 ns) beyond the available semiconducting APDs. The SNSPDs may contribute to the effective characterization of single-photon sources around a wavelength of 940 nm, which is promising for future QI processing. Further work to improve SDE and shorten the recovery time is ongoing.

**ACKNOWLEDGMENTS**

This work was funded by the NSFC (61401441 & 61401443), Strategic Priority Research Program (B) of the CAS (XDB04010200 & XDB04020100), 973 Program of China (2011CBA00202).

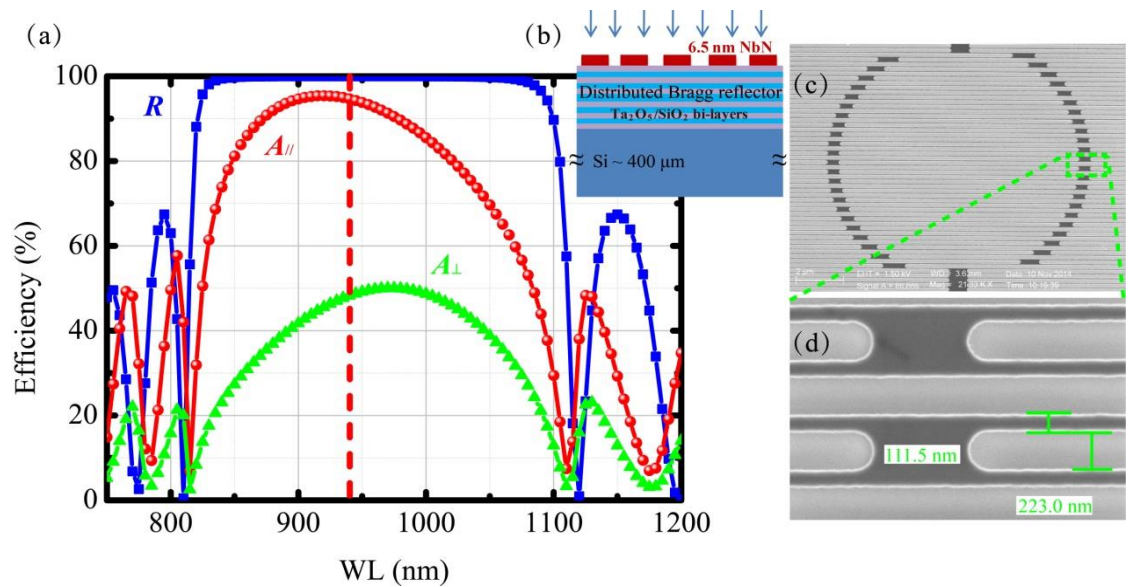

FIG. 1. (Color online) (a) Simulated reflectance, $R$, of DBR/Si substrate (blue squares) and absorptance, $A$, of the SNSPD. $A_{//}$ (red dots) and $A_{\perp}$ (green triangles) represent $A$ for parallel and perpendicular polarizations, respectively; (b) Schematic of SNSPD on DBR/Si substrate; (c) Scanning electron micrograph of an NbN SNSPD with an active area diameter of 10 μm; (d) Magnified image of the nanowire around the corner showing a well-defined linewidth.

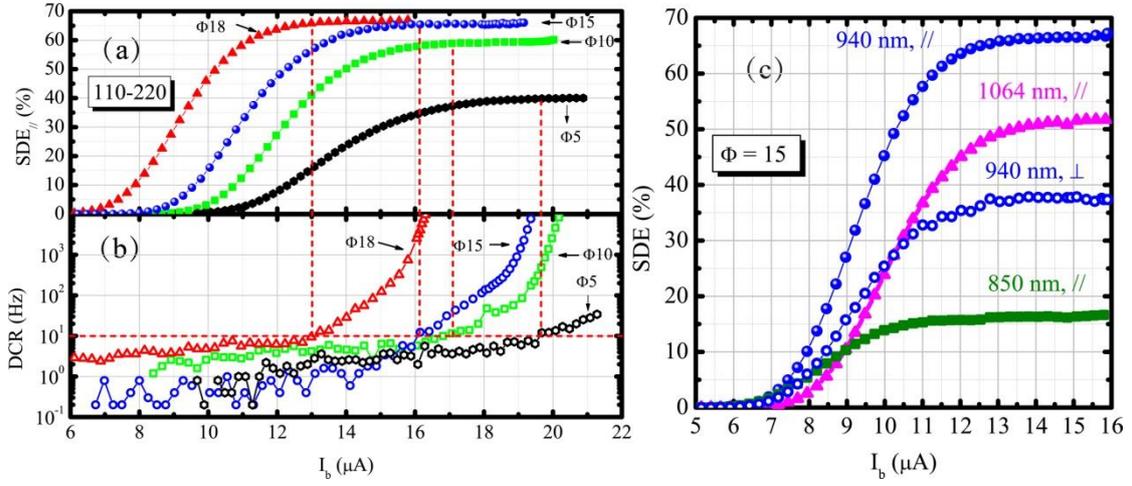

FIG. 2. (Color online) (a) Parallel-polarized SDE ($SDE_{//}$) vs. the bias current ($I_b$) for SNSPDs with various sizes at 2.2 K under illumination at a wavelength of 940 nm; (b) DCR vs. $I_b$. The dashed line indicates the DCR at 10 Hz; (c) $SDE_{//}$ at wavelengths of 850, 940, and 1064 nm and $SDE_\perp$ of 940 nm vs. $I_b$ for the device with $\Phi = 15$ μm.

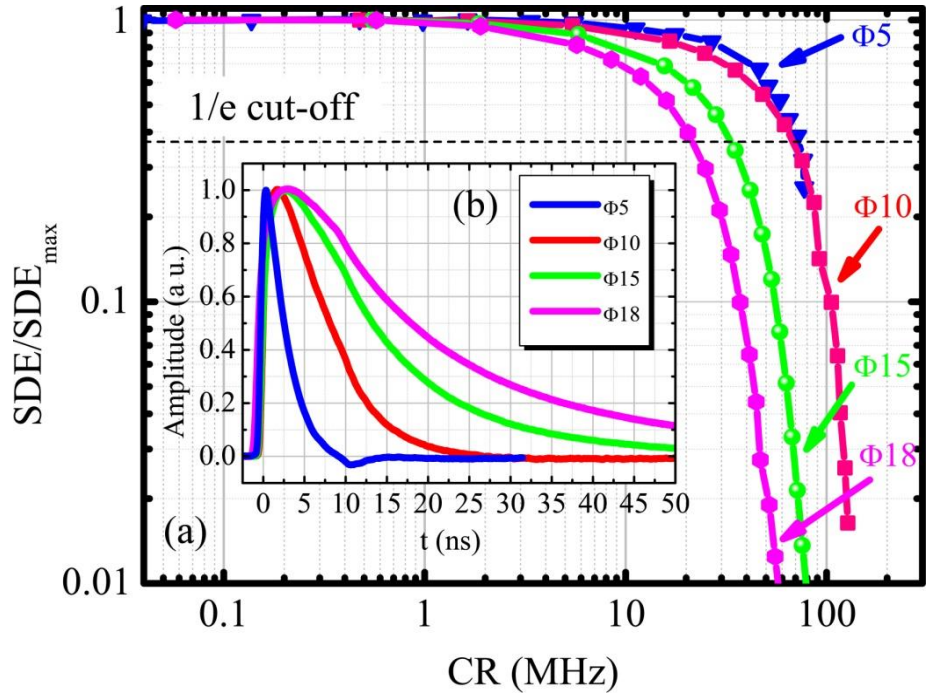

FIG. 3. (Color online) (a) Counting rate (CR) dependence of the normalized SDE for devices with $\Phi$ = 5, 10, 15, and 18 μm. CR was read out by the conventional circuit with a 50 Ω shunting resistor;(b) Averaged response pulse waveform.